\documentstyle[prl,aps,floats,epsf]{revtex}
\begin{document}
\twocolumn[\hsize\textwidth\columnwidth\hsize\csname@twocolumnfalse%
\endcsname

\draft

\title{\Large\bf Underlying Pairing States in Cuprate Superconductors}

\author{\bf Wei-Min Zhang} 

\address{Department of Physics, National Cheng Kung University, Tainan, 
		Taiwan 701, Republic of China}

\date{\today}
\maketitle

\begin{abstract}
{In this Letter, we develop a microscopic theory to describe 
the close proximity between the insulating antiferromagnetic (AF) 
order and the $d$-wave superconducting (dSC) order in cuprates. 
We show that the cuprate ground states form a configuration of 
coherent pairing states consisting of extended singlet Cooper 
pairs and triplet $\pi$ pairs, which can simultaneously 
describe AF and dSC orders.}
\end{abstract} 
%
\pacs{PACS numbers: 74.20.-z, 74.20.Mn, 74.25.Ha, 71.10.-w}
]

Discovery of high-$T_c$ superconductivity in copper-oxides
\cite{hightc} reveals very attractive, but also extremely 
complicated, new phenomena in strongly correlated electron 
systems\cite{Dag}. One of the most striking phenomena is 
the close proximity between the Mott insulating 
antiferromagnetic (AF) order and the $d$-wave superconducting 
(dSC) order. Theorists have tried various 
different mechanisms, such as the theory of resonating valence 
bond (RVB) states of singlet pairs introduced by Anderson a 
decade ago\cite{anderson87} and the SO(5) unified theory of 
AF and dSC order parameters proposed by S. C. Zhang
recently\cite{szhang97}, to explain this metal-insulator
transition. However, a microscopic description of the
AF to dSC transition has not yet been completed. 

It is believed that the underlying dynamics 
of strongly correlated electrons in cuprate superconductors 
can be modeled by the one-band 2D Hubband model or the $t-J$ 
model\cite{anderson87,fzhang88}. Unfortunately, the $t-J$ 
model itself is difficult to solve. 
Anderson's RVB state is a Mott insulating singlet at 
half-filling. However, it does not have AF ordering.  
In Zhang's SO(5) theory, AF order parameters are a part of 
the building blocks. However, near half-filling, this AF 
order may not be Mott insulating, due to the lack of
the constraint of no-doubly occupied sites which can destroy 
the SO(5) group structure.  Gauge theories later developed in 
terms of slave bosons in the charge spin separation scenario of 
RVB states\cite{Lee}, and various microscopic tests of the 
SO(5) theory made in the past year\cite{zhang98}, certainly 
provided deeper understanding to high $T_c$ superconductivity, 
but the mechanism concerning the AF to dSC transition remains 
unsolved.   
  
Most likely, the low energy degrees of freedom in 
cuprates should still be dominated by electron pairs, although 
pairing structures may be different from the singlet Cooper 
pairs in BCS theory or the triplet pairs in He$^3$ 
superfluid theory.  Experiments confirmed the existence of 
the $d_{x^2-y^2}$ pairing symmetry in cuprate superconductors. 
However, this does not imply that these electron pairs must  
all be spin singlet pairs, even though cuprate ground states 
are total spin singlet. Indeed, with $d$-wave singlet pairs 
alone, it is difficult to make a manifestation of the AF to
dSC transition.

In this Latter, we start with an insulating AF ordering 
of the undoped (half-filling) cuprate compound to 
explore how superconductivity emerges in doped cuprates. 
We find that insulating AF and dSC orders can be 
simultaneously described by a configuration of coherent 
pairing states consisting of extended singlet Cooper 
pairs and triplet $\pi$ pairs.  At half-filling, this 
configuration is dominated by the insulating AF ordering 
arisen from the mixing of singlet and triplet pairs. 
Upon doping with holes the $d$-wave pairing gap appears,
and above a certain level cuprates become superconducting. 
In this description, the nature of no-doubly 
occupied sites and the present of triplet 
$\pi$ pairs play an important role.

We shall begin with the formation of electron pairs in 
strongly correlated systems. The square lattice structure 
of layered copper-oxides allows us to restrict the problem  
in the reduced (half) first Brillouin zone in momentum space. 
Then the cuprate ground states may be obtained by projecting 
out all states of no-doubly occupied sites from the 
following generalized pairing state $|\Phi \rangle$: 
\begin{eqnarray}
	| \Phi \rangle &=& {\prod}'_{\bf k}\exp  \Big\{\eta_1({\bf k})
		c^\dagger_{{\bf k} \uparrow}c^\dagger_{-{\bf k}\downarrow} 
		+ \eta_2({\bf k})c^\dagger_{{\bf k}+{\bf Q} \uparrow}
		c^\dagger_{-{\bf k}+{\bf Q} \downarrow}  \nonumber \\
		& & ~~~~~~~~~ + \eta_3({\bf k})c^\dagger_{{\bf k} \uparrow}
		c^\dagger_{-{\bf k}+{\bf Q}\downarrow} + \eta_4({\bf k})
		c^\dagger_{{\bf k} \downarrow}c^\dagger_{-{\bf k}+{\bf Q}
		\uparrow} \nonumber \\
		& & ~~~~~~~~~ + \eta_5({\bf k})c^\dagger_{{\bf k} \uparrow}
		c^\dagger_{-{\bf k}+{\bf Q}\uparrow}+ \eta_6({\bf k})
		c^\dagger_{{\bf k} \downarrow}c^\dagger_{-{\bf k}+{\bf Q}
		\downarrow}  \nonumber \\
	& & ~~~~~~~~~- {\rm H.c} \Big\} | 0 \rangle .	\label{gcs}
\end{eqnarray}
where the complex parameters 
$\eta_i({\bf k})$ are generally {\it link-dependent} 
pairing wave functions. The parity symmetry in the $x-y$ plane 
requires $\eta_i({\bf k}) = \eta_i(-{\bf k})$. In the generalized 
coherent state theory \cite{wzhang90}, $|\Phi \rangle$ is a 
product (over  ${\bf k}$) of the local SO(8)/U(4) coherent pairing
states \cite{wzhang87}. 

In fact, Eq.~(\ref{gcs}) is also a multi-pair generalization 
of BCS pairing states or RVB states 
(after projection). Explicitly, we may rewrite 
$\eta_1({\bf k})=\eta_s({\bf k})+ \eta_d({\bf k})$, $\eta_2
({\bf k})=\eta_s({\bf k}) - \eta_d({\bf k})$, $\eta_3({\bf k})=
\eta_{\pi_0}({\bf k}) + \eta_\eta({\bf k})$, $\eta_4({\bf k})=
\eta_{\pi_0}({\bf k}) -\eta_{\eta}({\bf k})$, and $\eta_5({\bf k})
=\eta_{\pi_{+}}({\bf k})$, $\eta_6({\bf k})=\eta_{\pi_{-}}({\bf k})$. 
Then, Eq.~(\ref{gcs}) consists of all electron pairs concerned 
in the study of superconductivity. 
These are the ordinary $s$-wave Cooper pairs given by 
the $\eta_s({\bf k})$ term, the extended (i.e. valence
bond) singlet Cooper pairs by the $\eta_d({\bf k})$ terms 
[including the extended $s$-wave, $d$-wave and $s+id$ pairs 
in terms of $\eta_d({\bf k}) = \eta^{es}_{\bf k}\gamma({\bf k})$, 
$\eta^{d}_{\bf k}d({\bf k})$ and $\eta^{s+id}_{\bf k} (\cos k_x 
+i \cos k_y)$ respectively], the quasi-spin $\eta$ pairs by the 
$\eta_\eta ({\bf k})$ term which generate quasispin SU(2) 
transformations, the singlet $p$-wave pairs [if one lets 
$\eta_\eta({\bf k})=\eta^p_{\bf k}(\sin k_x \pm \sin k_y)]$, 
and finally the triplet $\pi$ pairs by $\eta_{\pi_i}({\bf k}) 
= \eta^{\pi_i}_{\bf k} d({\bf k})$, here $\gamma({\bf k})=
\cos k_x + \cos k_y$ and $d({\bf k})=\cos k_x - \cos k_y$. 

At half filling, without any loss of generality, 
we set $\eta_s({\bf k})=\eta_\eta ({\bf k})=0$, i.e., 
$\eta_1({\bf k})=-\eta_2({\bf k})$ and $\eta_3({\bf k})=
\eta_4({\bf k})$ for the AF ordered phase. In addition, 
the cuprate ground states must also be spin singlet 
due to the spin rotational symmetry which implies $\langle 
\Phi | S^\alpha |\Phi \rangle$ = 0. This can be satisfied 
if $\eta_5({\bf k})= - \eta_6({\bf k})$. 
These restrictions on pairing wave functions 
can be justified by explicitly calculating  
the AF order parameters from (\ref{gcs}):
\begin{eqnarray}
	m_z 
		&=& -{1\over 2N}{\sum}'_{\bf k} \Big[(z_{1\bf k}-
		z_{2\bf k})(z_{3\bf k}^*+z_{4\bf k}^*)
		+ {\rm H.c}\Big] , \nonumber \\
	m_+ &=& m_-={1\over 2N}{\sum}'_{\bf k} \Big[(z_{1\bf k}-
		z_{2\bf k})(z_{5\bf k}^*-z_{6\bf k}^*) 	
		+ {\rm H.c}\Big] , \label{sam}
\end{eqnarray}
where $m_z$ and $m_\pm$ are the average staggered magnetizations 
in the $z$ direction and the $x-y$ plane, respectively, $N$ is the 
total number of lattice sites, and the parameters $z_{i\bf k}$ are 
elements of the $4\times 4$ block matrix $Z_{\bf k}$ in the
canonical Bogoliubov transformations induced by (\ref{gcs}): 
$Z_{\bf k}=\eta \sin\sqrt{\eta^\dagger \eta}/\sqrt{\eta^\dagger 
\eta}$ with 
\begin{equation}
	\eta ({\bf k}) = {1\over 2} \left[\begin{array}{cccc}0 &
		\eta_1({\bf k}) 
		& \eta_5({\bf k}) & \eta_3({\bf k})\\ -\eta_1({\bf k}) 
	& 0 & \eta_4({\bf k}) & \eta_6({\bf k}) \\ -\eta_5({\bf k}) 
	& -\eta_4({\bf k}) & 0 & \eta_2({\bf k}) \\ -\eta_3({\bf k}) 
	& -\eta_6({\bf k}) & -\eta_2({\bf k}) & 0 \end{array} \right].
\end{equation}
The undoped ground state maximizes the staggered 
magnetization which corresponds to
\begin{equation}
	z_{1\bf k}=-z_{2\bf k}=z_{d\bf k}~,~~z_{3\bf k}=z_{4\bf k}~,
		~~z_{5\bf k}=-z_{6\bf k}.  \label{phs}
\end{equation}  
This is the same as letting $\eta_1({\bf k})=-\eta_2({\bf k})$, 
$\eta_3({\bf k})=\eta_4({\bf k})$ and $\eta_5({\bf k})= - 
\eta_6({\bf k})$. These restrictions on the pairing wave 
functions are also necessary for the manifestation of the extended 
(including $d$-wave) pairing symmetry of valence bonds, because 
Eq.~(\ref{phs}) implies $\eta_i({\bf k+Q}) = - \eta_i({\bf k})$.

In the above calculation, we have not imposed the constraint 
of no-doubly occupied sites. However, the result can be 
further varified under this constraint.  Here, instead of 
using Gutzwiller projector $P_G$ to remove doubly occupied 
sites from $|\Phi \rangle$, it is 
equivalent to require that $|\Phi \rangle$ must satisfy the 
constraint $\langle \Phi |\sum_i (n_{i\uparrow}n_{i\downarrow})| 
\Phi \rangle = 0$, where summation to $i$ is over the lattice 
sites.  This is because the operator $n_{i\uparrow}n_{i\downarrow}$ 
is positive definite. In the mean-field approximation of
(\ref{gcs}), we obtain the global constraint:
\begin{equation}
	{n^2 \over 4N^2} - m^2_s + |\Delta_s|^2
		+ |\Delta_\eta|^2 = 0 ,	\label{gcdo}
\end{equation}
where $n=\langle \Phi | \hat{n} | \Phi \rangle $ is the 
total electron number:
\begin{equation}
	n = 2 {\sum}'_{\bf k} {\sum}_{i=1}^6 |z_{i \bf k}|^2
		= N(1-\delta)   \label{pnc}
\end{equation}
and $\delta$ the fractional doping of holes, $m_s$ denotes the 
magnitude of the long-range AF order parameter determined 
from (\ref{sam}) by $m^2_s=m^2_z+m_+m_-$, and $\Delta_s$ 
and $\Delta_\eta$ are averaged order parameters of the 
ordinary $s$-wave Cooper pairs and the quasispin $\eta$ pairs 
which {\it vanish} under (\ref{phs}): $\Delta_s=\Delta_\eta
=0$. Then, Eq.~(\ref{gcdo}) reduces to 
\begin{equation}
	m^2_s = {n^2 \over 4N^2} .	\label{cndo}
\end{equation}
At half-filling ($\delta=0$), it gives $m_s = 0.5$ which 
is the same result as in N\'{e}el states. 
 
Furthermore, because of the spin rotational symmetry, 
without loss of generality we can define 
$z_{3\bf k}=z_{\pi\bf k}\cos 2\theta_{\bf k}$ and 
$z_{5\bf k}=z_{\pi\bf k}\sin 2\theta_{\bf k}$. 
Thus, from (\ref{cndo}), we find that $\theta_{\bf k} 
=\theta$ for all {\bf k} \cite{int1} and 
\begin{equation}
			z_{\pi\bf k} = z_{d \bf k}
	~~{\rm or}~~ z_{\pi\bf k} = - z_{d \bf k} . \label{scdo}
\end{equation}
This indicates that under the constraint of no-doubly 
occupied sites, the extended singlet Cooper pairs and 
triplet $\pi$ pairs in $|\Phi\rangle$ have {\it 
equal occupied probabilities}. Also, Eqs.~(\ref{sam}) 
and (\ref{pnc}) reduces to 
\begin{equation}
	 m_s = {4 \over N} {\sum}'_{\bf k} |z_{d\bf k}|^2
		~~,~~ n=8{\sum}'_{\bf k}|z_{d\bf k}|^2
		=N(1-\delta) \label{afo}
\end{equation}
and $m_z=m_s\cos 2\theta, m_+=m_-=m_s\sin 2\theta$.
Here $m_s$ and $n$ can be written only in terms 
of the singlet pairing wave function $z_{d\bf k}$ 
because of the constraint of no-doubly occupied 
sites (\ref{scdo}). At half-filling, 
electrons are uniformly distributed. Then 
Eq.~(\ref{afo}) gives 
\begin{equation}
	|z_{d\bf k}| = 1/2  .  \label{lcdo}
\end{equation}
This is indeed the local constraint of no-doubly occupied 
sites at half-filling. One can show that $\langle \Phi 
|(\hat{n}-n)^2|\Phi \rangle =0$ under (\ref{scdo}) and 
(\ref{lcdo}).

The above rigorous results are derived only from the 
kinematics of the coherent pairing state 
$|\Phi \rangle$. Dynamically, we may determine the 
ground state from the $t-J$ model. The $t$-term 
is a hoping Hamiltonian $H_t = \sum_{{\bf k},\sigma} 
\varepsilon({\bf k}) c^\dagger_{k\sigma} c_{k\sigma}$, 
where $\varepsilon({\bf k})=-2t (\cos k_x + \cos k_y)$. 
Its expectation value is 
\begin{equation}
	\langle H_t \rangle = 2{\sum}'_{\bf k} \varepsilon({\bf k})
		(|z_{1\bf k}|^2 - |z_{2\bf k}|^2)  \label{tterm}
\end{equation}
which vanishes by (\ref{phs}), as expected from the 
constraint of no-doubly occupied sites at half-filling. 
Under the restriction (\ref{phs}), the expectation 
value of the $J$-term [$H_J=J\sum_{\langle i,j\rangle} 
(S_i \cdot S_j - {1\over 4}n_i n_j )$] is given by
\begin{equation}
	{\langle H_J \rangle\over JN} = -\Big(|\Delta_d|^2 + 
		|\Delta_{es}|^2 + 2m_s^2 +{n^2\over 2N^2} \Big) ,
		 \label{hmh}
\end{equation}
where the averaged $d$-wave pairing order parameter  
\begin{eqnarray}
	\Delta_d &=& {1 \over N} {\sum}'_k d({\bf k}) \Big(z^+_{\bf k}
		w^+_{\bf k} +  z^-_{\bf k}w^-_{\bf k} \Big) \nonumber \\
	&=& {2 \over N} {\sum}'_{\bf k} d({\bf k}) 
		z_{d\bf k} \sqrt{1-4|z_{d\bf k}|^2} , \label{dgap}
\end{eqnarray}
and the extended $s$-wave pairing order parameter $\Delta_{es}$ 
has the same form as (\ref{dgap}) by replacing $d({\bf k})$ 
with $\gamma({\bf k})$. 
The parameters $z^\pm_{\bf k}=z_{d\bf k} \pm z_{\pi\bf k}$ and 
$w^\pm_{\bf k}=\sqrt{1-|z^\pm_{\bf k}|^2}$. Note that although 
it can be written in terms of $z_{d\bf k}$ in the second 
equality of (\ref{dgap}) [because of the constraint (\ref{scdo})],
$\Delta_d$ receives contributions from both the singlet 
and triplet pairs. As a result, $\Delta_d$ behaves very 
different from that in pure singlet pair states. At
half-filling, (\ref{lcdo})  
ensures $\Delta_d=\Delta_{es}=0$. Then $\langle H_J 
\rangle=-JN$ which again gives the same result as in N\'{e}el 
states.  

These results are very remarkable. It shows that  
under constraints (\ref{phs}) and (\ref{scdo}),
the coherent pairing state $|\Phi\rangle$ is 
insulating AF ordering at half-filling. It consists
of extended singlet Cooper pairs and triplet 
$\pi$ pairs with equal pair numbers. The singlet
pairs are indeed Anderson's RVB pairs \cite{anderson87}, 
while the triplet $\pi$ pairs are the $\pi$ operators 
of Demler and Zhang \cite{szhang95}. Both the AF and 
dSC order parameters are functions of extended 
pairing wave functions (including the extended $s$-wave, 
$d$-wave and $s+id$ pairing functions\cite{dwave}) 
of singlets and triplets. However, because of
the constraint of no-doubly occupied sites, the 
pairing order parameter vanishes while the AF 
order parameter is the same as in N\'{e}el states. 

Meanwhile, although the configuration of coherent 
pairing states contains the same AF properties as the 
N\'{e}el configuration, these two configurations 
are different. Electrons in N\'{e}el states are 
not paired and N\'{e}el configuration itself is only 
valid for half-filling. However,
$|\Phi\rangle$ is defined for any doping and it
is built with extended (valence bond) pairs. Thus,
$|\Phi\rangle$ provides a natural path to explore
how cuprates undergo the transition from an insulating 
AF order to a dSC order after dopings.
 
Upon doping with holes, it is commonly believed that  
hopings rapidly distroy the AF ordering. 
Eq.~({\ref{tterm}) shows that the lowest energy 
gained from the $t$-hoping corresponds 
to $z_{2\bf k}=0$, which reduces AF 
order parameter by a half, as shown by (\ref{sam}).  
Apparently this is a nice result as expected from
hoping dynamics. However, $z_{2\bf k}=0$ leads to  
a non-vanishing $\Delta_s$. A non-vanishing $\Delta_s$ 
enhances the AF order parameter $m_s$ by the constraint 
(\ref{gcdo}), which is contrary to (\ref{sam}) for 
$z_{2\bf k}=0$. Indeed, the ordinary Cooper pairs are 
on-site pairs which are excluded by the constraint 
of no-doubly occupied sites. In terms of energy scales, 
the on-site repulsive Coulomb interaction is much stronger 
than hopings, a non-vanishing $\Delta_s$ contributes 
a large positive Coulomb energy to the ground states, 
which is certainly unfavored. Indeed, any net contribution 
from the $t$-hoping in mean-field theory requires 
$|z_{1\bf k}|^2 \neq |z_{2\bf k}|^2$, which
will break the extended singlet pairing symmetry. 
Hence, doped ground states must still be 
restricted by (\ref{phs}). 
 
As a result, after dopings, $|\Phi \rangle$ still consists 
of the extended singlet Cooper pairs and triplet $\pi$ pairs. 
All equations derived above are still valid except for 
(\ref{lcdo}). Eq.~(\ref{afo}) shows that the 
AF order parameter $m_s$ is decreased linearly with doping 
$\delta$. By further including the quantum fluctuation of 
spin-wave excitations, $m_s$ is suppressed from 0.5 to 0.3 at 
half-filling \cite{Henly}, and will be suppressed
as well upon dopings. While, the pairing order 
parameter $\Delta_d$ [given by Eq.~(\ref{dgap})] is zero 
at half-filling. Then it is increased after dopings
so that superconducting states will emerge at a certain 
level of dopings. 

To be more explicit, we may take again the $t-J$ 
model as an example. When all electrons in ground states 
are paired, the $t$-term vanishes by the pairing 
symmetry $z_{1\bf k}=-z_{2\bf k}$ for any doing.
However, the next order hoping ($t'$-term) has a non-zero
expectation value in $|\Phi\rangle$. The ground states 
can be determined by minimizing the $t'-J$ Hamiltonian, 
which leads to $z_{d\bf k} \rightarrow |z_{d\bf k}|e^{i\phi}$ 
and the gap equation at zero-temperature,
\begin{equation} \label{gapeq}
	\Delta_{\bf k} = {1 \over N}{\sum}'_{\bf k}V_{\bf kk'}
	{\Delta_{\bf k'} \over 2E_{\bf k'}} ,
\end{equation}
where $\Delta_{\bf k}\equiv \Delta_d d({\bf k}) + \Delta_{es}
\gamma({\bf k})$, 
$V_{\bf kk'}=J[d({\bf k})d({\bf k'})+ \gamma({\bf k})
\gamma({\bf k'})]$ and $E_{\bf k}=\{J^2\Delta^2_{\bf k}+
[\varepsilon({\bf k})-\mu-2(1-\delta)J]^2\}^{1/2}$ with 
$\varepsilon({\bf k})=-4t'\cos k_x\cos k_y$. The fixed 
electron number gives $-{2\over N}{\sum}'_{\bf k}
{\varepsilon({\bf k})-\mu - 2(1-\delta)J \over 
E_{\bf k}}=1-2\delta$, and $\mu$ is the chemical 
potential.

The solutions of (\ref{gapeq}) show that 
the $d$-wave gap order parameter appears after dopings 
for $\delta <0.5$ with maximum peak
$\Delta_d \simeq 0.07 \sim 0.10$ at $\delta \simeq 0.15 
\sim 0.20$ (the typical optimal doping region)
for $t'/J=0.30 \sim 0.20$. This is in good agreement 
with the experimental observations of $d$-wave 
superconducting states in cuprates. It is also 
striking that the extended $s$-wave superconducting
states only emerge in overdoped region of $\delta > 0.5$.
The separation of the $d$-wave states in optimal 
dopings from the extended $s$-wave states in  
overdopings is controlled by the $t'$-term. Here $t'$ must 
be positive. For a negative $t'$, the ordering of $\Delta_d$ 
and $\Delta_{es}$ in terms of $\delta$ will be exchanged.
If we let $t'=0$, then $\Delta_d 
= \Delta_{es}$ which have the maximum value 
at doping $\delta=0.5$. This corresponds to the 
symmetry limit of Zhang's SO(5) theory [see (\ref{hmh})],
although generally the pairing wave functions determined 
here do not form a rigorous global SO(5) group 
structure \cite{Henley}. The numerical results 
are plotted in Fig.~1.  
\begin{figure}
  \begin{center}
	\input{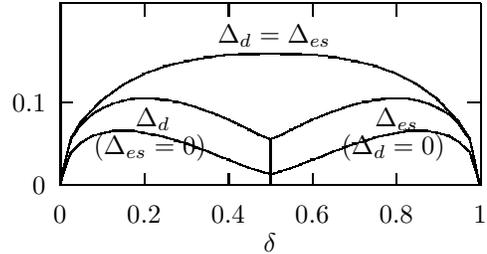}
  \end{center}
  \vskip -0.15in
  \caption{Gap order parameters via dopings at 
zero temperature. The (top) solid line is for 
$t'/J=0$, and the middle and bottom lines 
are for $t'/J=0.2$ and 0.3, respectively.}
\end{figure}

The above results are very different from the $d$-wave
singlet pair theory. The $d$-wave singlet pair states
correspond to the special case of (\ref{gcs}) with
$\eta_1({\bf k})=-\eta_2({\bf k})=\eta_d{(\bf k})$ and other 
$\eta_i({\bf k})=0~ (i=$3 to 6). Then from (\ref{sam}),
(\ref{pnc}) and (\ref{dgap}), we have
\begin{eqnarray}
	& & m_s=0~~,~~ n=4{\sum}'_{\bf k}|z_{d\bf k}|^2
		=N(1-\delta) \nonumber \\	
	& & \Delta_d = {2 \over N} {\sum}'_{\bf k} d({\bf k}) 
		z_{d\bf k} \sqrt{1-|z_{d\bf k}|^2} . \label{agap}
\end{eqnarray}
Hence, with singlet pairs alone, no AF order exists. 
The order parameter $\Delta_d$ has a maximun value near 
half-filling ($z_{1\bf k} \simeq \sqrt{1/2}$), then it is 
slowly decreased with dopings \cite{dwave}. Of course, 
here the constraint of no-doubly occupied sites cannot 
been imposed analytically by (\ref{cndo}) so that 
auxiliary-boson variables must be introduced\cite{Lee}.

Formally, (\ref{agap}) and (\ref{dgap}) show that the 
magnitude of $\Delta_d$ in pure singlet pair states is 
about twice of that in mixed states of singlet and triplet 
pairs. In other words, pure singlet pair states have 
lower gap energies than mixed pair states. However,
mixed pair states maximize the AF energy 
which cannot be taken into account in pure singlet 
pair states [see Eq.~(\ref{hmh}) in the $t-J$ model, for
an example]. Because of the AF energy, mixed pair 
states have indeed a lower ground state energy than 
pure singlet pair states in the mean-field theory.  

Summarizing the above discussions, we show that 
underlying pairing states undergoing an insulting 
AF to dSC transition can be determined by the coherent 
state $|\Phi\rangle$ of extended (valence bond) 
singlet Cooper pairs mixed with triplet $\pi$ pairs. 
These coherent pairing states are unconventional, and 
can be regarded as a nontrivial generalization of 
Anderson's singlet pair RVB states.
Advantage of the coherent pairing state 
$|\Phi \rangle$ is that it contains an intrinsic AF order 
which is Mott insulating at and near half-filling, while
in optimal dopings, it has the $d$-wave SC order. 
Therefore, it can {\it microscopically} unify the 
insulating AF and dSC orders in the same configuration. 
This is for the first time to analytically show that a
pairing state can simultaneously describe both the AF 
order and the dSC order. 
Dynamically, we show that due to the extended singlet 
pairing symmetry, the contribution from the $t$-hoping is, 
surprisingly, cancelled in (classical) ground states. 
The exchange interaction ($J$-term) manifests a symmetry 
limit of Zhang's SO(5) theory.
While, the insulating AF to dSC transition under dopings
that is comparable with experimental observations 
can be determined by the next nearest-neighbor hopings 
(with a positive but weak $t'$ hoping). 

Furthermore, the canonical transformation structure of 
(\ref{gcs}) generates a quasiparticle picture, 
which allows us to easily determine 
normal state properties of cuprates as well as the 
phase diagram of high $T_c$ superconductivity in the
$T-\delta$ plane  \cite{wzhang99b}. 
Also note that cuprate ground 
states determined here by minimization are  
classical states. In these classical 
ground states, $z_{d\bf k} \rightarrow 
|z_{d\bf k}|e^{i\phi}$ and $\theta_{\bf k} \rightarrow 
\theta$ so that the configuration manifests the global 
SU$_{\rm spin}$(2)$\times$U$_{\rm gauge}$(1) degeneracy. 
This SU(2)$\times$U(1) symmetry will be spontaneously
broken at quantum level. The quantum fluctuations of 
$z_{d\bf k}$ and $\theta_{\bf k}$ describe pairing 
excitations and spin-wave excitations (Goldstone modes)
which further lower energies of the ground states over 
all dopings. Therefore, these fluctuations play a 
very important role in the determination of dynamical 
properties of electrons in cuprates. A typical evidence
is the AF magnetic ordering which is too strong 
in classical ground states [see (\ref{cndo})]. 
However, it can be largely suppressed (and should be 
cancelled in optimal dopings) by the quantum 
fluctuation of spin-wave excitations \cite{Henly}. 
In separate publications \cite{wzhang99c}, 
based on the generalized coherent state theory\cite{wzhang90}
we shall use $|\Phi\rangle$ to develop a theory of 
quantum fluctuations to detail these features, and then 
to describe dynamical and thermal properties of high 
$T_c$ superconductivity in cuprates. 
  
I wish to thank T. K. Lee, J. X. Li, Q. Niu and C. S. Ting
for many useful discussions.



\begin{references}
\bibitem{int1} This is a manifestation of the global spin rotational 
	symmetry. Specifying the value of $\theta$ spontaneously breaks 
	the spin rotational symmetry, which leads to the gapless spin-wave 
	excitations.
\end{references}
\end{document}